\documentclass{article}

\usepackage{PRIMEarxiv}

\usepackage[utf8]{inputenc} 
\usepackage[T1]{fontenc}    
\usepackage{hyperref}       
\usepackage{url}            
\usepackage{booktabs}       
\usepackage{amsfonts}       
\usepackage{nicefrac}       
\usepackage{microtype}      
\usepackage{lipsum}
\usepackage{fancyhdr}       
\usepackage{graphicx}       
\graphicspath{{media/}}     
\usepackage{amsmath}
\usepackage{xcolor}
\usepackage{soul}

\newcommand{\tb}[1]{\textcolor{blue}{#1}}

\title{DiffNMR2: NMR Guided Sampling Acquisition Through Diffusion Model Uncertainty
}

\author{
  Etienne Goffinet, Sen Yan *, Fabrizio Gabellieri, \\
  Laurence Jennings, Lydia Gkoura, Filippo Castiglione, \\
  Ryan Young, Idir Malki, Ankita Singh,
  Thomas Launey\\
  \\
  Biotechnology Research Center\\ 
  Technology Innovation Institute\\ Abu Dhabi,
  UAE\\
  \\
  \texttt{yansen0508@gmail.com} \\
}

\begin{document}
\maketitle

\begin{abstract}
Nuclear Magnetic Resonance (NMR) spectrometry uses electro-frequency pulses to probe the resonance of a compound's nucleus, which is then analyzed to determine its structure. The acquisition time of high-resolution NMR spectra remains a significant bottleneck, especially for complex biological samples such as proteins. In this study, we propose a novel and efficient sub-sampling strategy based on a diffusion model trained on protein NMR data. Our method iteratively reconstructs under-sampled spectra while using model uncertainty to guide subsequent sampling, significantly reducing acquisition time. Compared to state-of-the-art strategies, our approach improves reconstruction accuracy by 52.9\%, reduces hallucinated peaks by 55.6\%, and requires 60\% less time in complex NMR experiments. This advancement holds promise for many applications, from drug discovery to materials science, where rapid and high-resolution spectral analysis is critical.
\end{abstract}

\keywords{Diffusion Model, Nuclear Magnetic Resonance, Uncertainty}


\begin{figure*}[tb]
    \centering
    \includegraphics[width=1\linewidth]{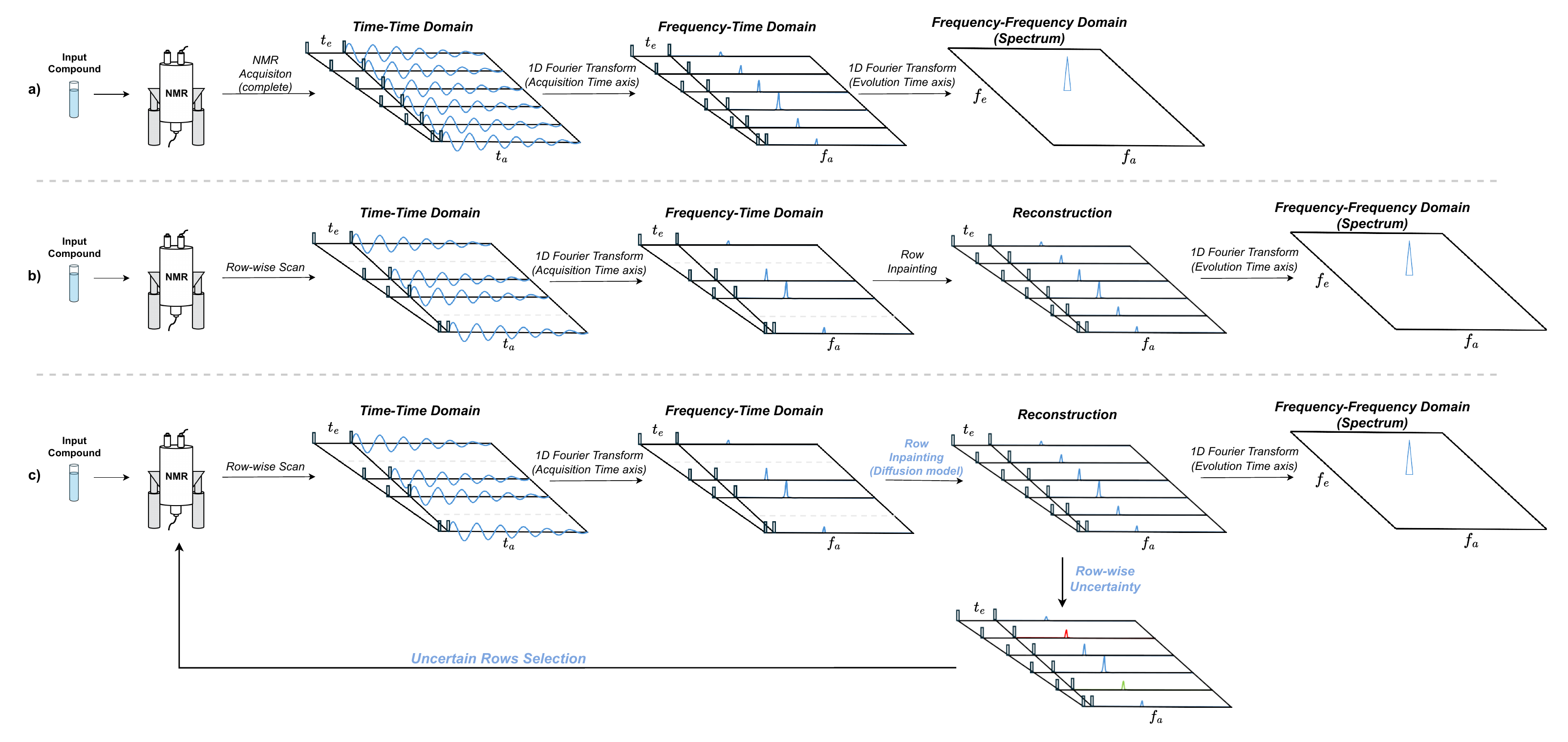}
    \vspace{-1cm}
    \caption{\textbf{a) Regular NMR experiment.}
    A 2D NMR resonance map in the time-time domain $(t_e,t_a)$ is produced by recording the response of nuclei subjected to two-pulse sequences, where each row is obtained with a different time delay (called evolution time: $t_e$) between the two pulses and $t_a$ represent the acquisition time. This map is first transformed into a frequency-time representation via a 1D Fourier Transform along the acquisition time axis: $(t_e,t_a) \to (t_e,f_a)$. A second 1D Fourier Transform along the evolution time axis produces the final spectrum: $(t_e,f_a)\to(f_e f_a)$.
    \textbf{b) Non-uniform sampling (NUS)}. Differing from the regular NMR experiment, Non-Uniform Sampling selects a subset of the evolution times (Row-wise Scan). After 1D Fourier Transform on the acquisition time axis, existing methods such as Compressed Sensing \cite{kazimierczuk2011accelerated} and Low-Rank Approximation \cite{qu2015accelerated} are applied to reconstruct the frequency-time domain (Row Inpainting). 
    \textbf{c) Guided sampling workflow.} Differing from NUS, we propose a novel dynamic sampling strategy by leveraging the diffusion model uncertainty. This is the first attempt using the diffusion model during NMR acquisition. The novelties are highlighted in blue texts.}
    \label{fig:nus_all}
\end{figure*}

\section{Introduction}
Nuclear Magnetic Resonance (NMR) spectroscopy is a powerful tool for identifying the molecular structures, providing insights into the arrangement of atoms within complex molecules. 
It has become an indispensable tool in chemistry, biochemistry, and structural biology from its introduction in 1938 \cite{giunta2020discovery}. 
However, acquiring NMR spectra requires long acquisition times 
to reconstruct high-resolution information (Fig.~\ref{fig:nus_all}). 
Traditional acquisition methods can be prohibitively slow in applications involving large biomolecules 
or when time is bounded, such as in drug discovery \cite{pellecchia2002nmr,pellecchia2008perspectives} or metabolomics \cite{markley2017future,emwas2019nmr}. 
For example, acquiring a 4D NMR spectrum (CCNOESY) of the phycobilisome linker polypeptide domain of CpcC takes up to 90 hours \cite{2l8v}.

Non-uniform sampling (NUS) addresses this challenge by randomly sub-sampling the scanned signal while enabling high-quality spectral reconstructions (Fig.~\ref{fig:nus_all}). 
Instead of sampling every data point in a regular pattern, NUS collects a subset of data points based on a pseudo-random or optimized sampling schedule \cite{marvasti1993nonuniform}.
The advantage of NUS lies in its capacity to significantly accelerate acquisition, but its effectiveness relies on robust reconstruction algorithms capable of handling incomplete data. 
Several methods have been developed for reconstructing spectra from NUS data, 
including Low-Rank Approximation (LR) \cite{qu2015accelerated}, 
Compressed Sensing (CS) \cite{kazimierczuk2011accelerated}, and 
Deep Learning (DL)-based methods \cite{qu2020accelerated,zhan2024fast,zheng2022fast,karunanithy2021fid}.

In this study, we introduce an AI-based approach that dynamically adapts the acquisition strategy. 
By integrating acquisition and reconstruction, the method leverages the uncertainty of the diffusion model 
(i.e., the degree of variability in the model's reconstruction)
to refine sampling in real-time, significantly reducing the acquisition burden. 
Using a real-life protein 2D NMR dataset \cite{klukowski2024100}, we demonstrate that our method outperforms state-of-the-art sampling techniques and reconstruction methods, delivering accurate high-resolution spectral reconstructions from substantially reduced datasets. 
This work establishes a new paradigm in NMR spectroscopy by combining the strengths of NUS and advanced machine learning, enabling rapid, high-quality spectral analysis for time-sensitive applications.

\section{2D NMR Data}
Two-dimensional nuclear magnetic resonance spectroscopy provides detailed structural and dynamic
information about molecules by resolving spectral data along two frequency axes. 
This is achieved by extending the principles of 1D NMR with the additional time domain, offering insights into interactions between nuclei.

\subsection{The Acquisition Process}



A 2D NMR experiment consists of three primary stages: preparation, evolution, and acquisition. During the preparation phase, tailored radio-frequency pulse sequences excite the nuclear spins and establish coherence pathways to encode the desired interactions.
The preparation stage is followed by the evolution phase, where nuclear spin information is read during an evolution time ($t_e$),
Revealing the chemical environment, such as chemical shifts and spin-spin couplings. 
The encoded magnetization is subsequently recorded during the acquisition time ($t_a$), generating dual time-domain data ($t_e$, $t_a$). 
These data are processed to produce a 2D frequency spectrum, where the indirect dimension ($t_e$) typically reflects spin-spin interactions, and the direct dimension ($t_a$) corresponds to the resonance frequencies of individual nuclei.

Pulse sequences are carefully designed radiofrequency pulses and delays that manipulate spin states to achieve specific coherence transfers and magnetization pathways.
For example, a simple COSY (Correlation Spectroscopy) experiment uses a two-pulse sequence to correlate chemical shifts of coupled nuclei, highlighting scalar (J) coupling. 
More complex sequences like HSQC (Heteronuclear Single Quantum Coherence) include additional steps, such as heteronuclear decoupling, to correlate proton and heteronucleus 
(e.g., $^{13}$C or $^{15}$N) chemical shifts. 
These advanced techniques are particularly valuable for analyzing larger biomolecules.
Finally, to enhance the signal-to-noise ratio, it is common practice to perform multiple scans of the same spectrum and average them to obtain the final resonance map.

\subsection{Data Processing}
The raw data from a 2D NMR experiment is initially represented as a 2D array of signal intensities varying over $t_e$ and $t_a$. This data is transformed into a frequency domain using a two-dimensional Fourier transform (see Fig.~\ref{fig:nus_all} (a). 
Before the transform, techniques like phase correction and apodization are applied to improve spectral resolution and signal-to-noise ratio. The resulting 2D spectrum is plotted on the frequency axes of the evolution axis and acquisition axis, with cross-peaks indicating interactions between nuclei.

\subsection{Non-Uniform Sampling (NUS)}
Non-Uniform Sampling (NUS) is a technique used to improve the efficiency of 2D NMR data acquisition by selectively sampling a subset of the evolution times ($t_e$) rather than collecting the full Nyquist grid.
Instead of acquiring data at every increment of $t_e$, only a fraction of the points are sampled at random. One standard strategy is the Sine-Weighted Poisson-Gap sampling \cite{hyberts2010poisson}, which assumes a Poisson distribution for the gap size between evolution times. 

The reconstruction of the full spectrum from sparsely sampled data requires advanced computational algorithms. One prominent method is \textbf{Low-Rank Approximation} \tb{(LRA)} \cite{qu2015accelerated}, which represents the data as a Hankel matrix and approximates it with a low-rank reconstruction that best fits the sampled points. 
While computationally efficient, the LRA struggles to capture highly complex spectral features or systems with noise, leading to artifacts in the reconstruction. \textbf{Compressed Sensing} \cite{kazimierczuk2011accelerated} is another standard that leverages the sparsity of NMR signals in the time-time domain to reconstruct the complete dataset. However, CS requires careful tuning of regularization parameters and may exhibit limitations in reconstructing spectra with dense peaks or overlapping signals.

In recent literature, Qu et al. \cite{qu2020accelerated} introduced a dense Convolutional Neural Network (CNN) \cite{lecun2015deep} with a spectrum consistency block for Poisson-Gap NUS data, achieving promising results for 2D and 3D spectra. 
Zhan et al. \cite{zhan2024fast} employed SEPSNet, an attention-based network, to enhance pure shift NMR reconstruction from a 2D resonance map. 
Zheng et al. \cite{zheng2022fast} developed PS-ResNet, a ResNet-based model, for denoising 1D spectra. 
Karunanithy et al. \cite{karunanithy2021fid} applied dilated convolutions through Fidnet, inspired by WaveNet, for synthetic NMR data analysis.
Despite these advancements, these deep learning methods share a common limitation: they are primarily trained on synthetic datasets, which may not fully capture the complexity and variability of real-life NMR data. 
As a result, their performance has not consistently surpassed the established CS baseline for NUS reconstruction, particularly when applied to experimental datasets \cite{yan2024diffnmr}.

\section{Proposed Method: Guided Sampling}
In contrast to existing methods, our approach employs a diffusion model in the frequency-time domain to reconstruct spectra with high precision, offering a novel perspective on NUS data reconstruction. By leveraging the uncertainty of the diffusion model in the reconstructed spectrum, we introduce an adaptive sampling strategy that dynamically prioritizes the evolution times  iteratively. 
This method enhances data acquisition efficiency, reduces overall acquisition time, 
and ensures high-quality spectral reconstructions, providing a practical and effective 
solution for experimental NMR data.

\subsection{NUS Reconstruction with Diffusion Model inpainting}
Denoising Diffusion Probabilistic Models (DDPM) \cite{ho2020denoising}, are a class of neural networks originally inspired by non-equilibrium thermodynamics processes \cite{sohl2015deep}. These models are trained by progressively destroying data with noise and then learning to reverse the process and regenerate the original data. 
The first phase (i.e., the diffusion process) consists of generating noisy data such that:
\begin{equation*}
    q(\mathbf{x}_t | \mathbf{x}_{t-1}) = \mathcal{N}(\mathbf{x}_t; \sqrt{1 - \beta_t} \mathbf{x}_{t-1}, \beta_t \mathbf{I}),
\end{equation*}
where $\mathbf{x}_0$ represents the original spectrum, $\mathbf{x}_t$ is the noisy spectrum at time step $t$, and $\beta_t$ is a variance parameter term controlling the amount of noise added at each step.

The second phase (i.e., denoising process), involves training a UNet \cite{ronneberger2015u} to minimize the difference between the predicted noise and the actual noise added in the forward process. 
This can be formulated as optimizing the Mean Reconstruction Error loss:

\begin{equation*}
L = \| \epsilon - \epsilon_\theta(\mathbf{x}_t, t) \|^2
\end{equation*}

\begin{figure}[tb]
    \centering
    \includegraphics[width=\linewidth]{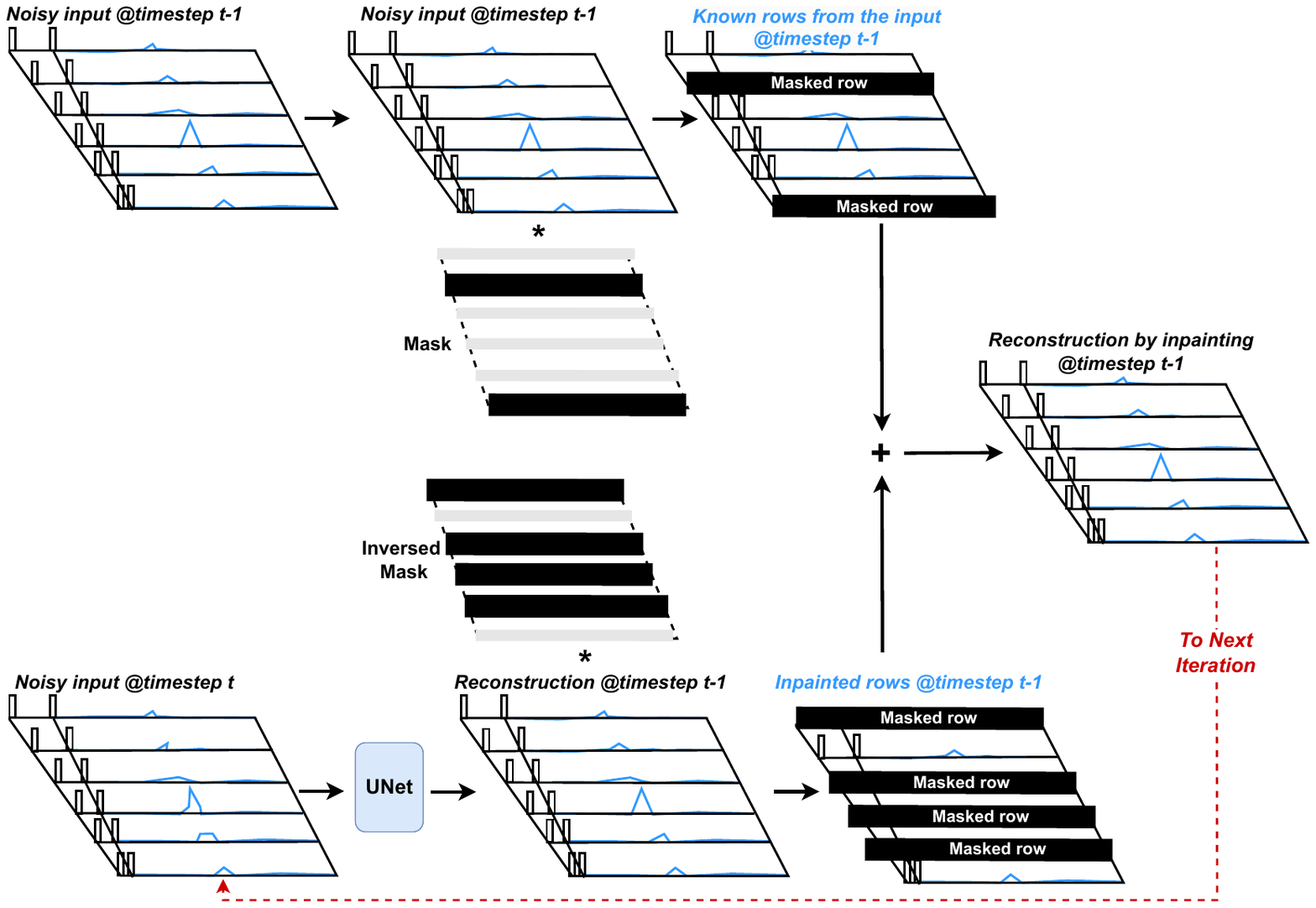}
    \caption{Repaint pipeline for the reconstruction of NMR data. 
    At variance with the conventional denoising process, in each step, we sample the known rows (top) from the input and reconstruct the inpainted rows from the UNet output (bottom).}
    \label{fig:inpainting}
\end{figure}

The same model and architecture can also be used for \emph{inpainting} in the context of incomplete image reconstruction. 
In our experiments, we perform the inpainting using a diffusion model trained for denoising combined with the \emph{Repaint} pipeline \cite{lugmayr2022repaint}. 
This pipeline works by combining the information of the unmasked area during the denoising step. This is  illustrated in Fig.~\ref{fig:inpainting}.

\subsection{Guided Sampling Workflow} 
Instead of revealing all evolution times of the frequency-time input as in regular NMR experiments (i.e., applying the whole pulse sequence) or scanning a random selection of evolution times as NUS (i.e., selecting a subset of input rows), Guided Sampling is an iterative approach that, at each step, prioritizes evolution times (input rows) to be scanned by the NMR machine.

Our proposed Guided Sampling approach is illustrated in Fig.~\ref{fig:nus_all} (c).
\textbf{Row-wise Scan.} At each iteration, a small percentage (e.g., 2\%) of the input rows is revealed (unmasked) by the NMR machine.
\textbf{1D Fourier Transform.} After performing a 1D Fourier Transform along the acquisition time axis, the data domain switches from Time-Time to Frequency-Time.
\textbf{Row Inpainting.} The diffusion model predicts the masked parts of this map by generating and averaging multiple outputs. 
\textbf{Row-wise Uncertainty.} Alongside the average, the pixel-wise uncertainty is calculated using the variance in the reconstructed map. For a given evolution time $e$ and frequency $f$, the pixel-wise variance is defined as:
\[
\sigma^2_{f,e} = \frac{1}{n-1} \sum_{i=1}^n \left(x_{f,e}^{(i)} - \mu_{f,e}\right)^2
\]
where $n$ is the number of reconstructed samples, $x_{f,e}^{(i)}$ is the pixel-wise value of the $i$-th sample, and $\mu_{f,e}$ is the mean pixel-wise value. 
The row-wise uncertainty for an evolution time $e$ is computed as the sum of the pixel-wise uncertainties:
\[
U_e = \sum_{f=1}^{f_a} \sigma^2_{f,e}
\]
where $f_a$ represents the total number of frequency points.
\textbf{Uncertain Rows Selection.} The rows with the highest uncertainty are then selected and scanned by the NMR machine in the next iteration. For initialization, the initial subset of input rows is selected using Poisson-Gap sampling, as row-wise uncertainty is not yet available.

\subsection{Data}
Recent advancements in diffusion models for image generation rely on massive datasets containing hundreds of millions of images \cite{schuhmann2021laion,rombach2022high}. 
In contrast, large-scale datasets for 2D NMR protein spectra are scarce, with only a few public repositories available. The standard approach has been to train models on synthetic spectra \cite{zhan2024fast,qu2020accelerated,karunanithy2021fid, zheng2022fast}, which often fail to capture the complex noise patterns and signal variations present in real-life data. This limitation also explains why Compressed Sensing \cite{kazimierczuk2011accelerated} and Low-Rank Approximation \cite{qu2015accelerated} remain the state-of-the-art methods for reconstructing real-world NMR spectra. To address this gap, our study focuses on training models exclusively on real-life NMR spectra, providing a more realistic and reliable basis for evaluation.

In this study, we use the 100-protein NMR spectra dataset \cite{klukowski2024100}, comprising 1329 spectra in 2D, 3D, and 4D formats. These spectra, derived from 100 proteins, were sampled using NMR machines operating at frequencies ranging from 600 to 950 MHz.
We augmented the dataset by including 2D spectra obtained by projecting higher-dimensional (3D or 4D) spectra into 2D representations. This technique, commonly used in NMR workflows for visualization purposes, increases the total number of samples to over 3500. The process maintains the integrity of the original signals while significantly enriching the dataset for training. Finally, the spectra are transformed with a 1D Inverse Fourier Transform to obtain their representation in the frequency-time domain where each row corresponds to a 1D NMR spectrum, and to an evolution time.

\subsection{Training} 
We split the dataset into training, validation, and test sets with respective proportions of 80\%, 10\%, and 10\%, ensuring the test set independence and avoiding data leakage. 
The diffusion model is trained by optimizing the Mean Reconstruction Error (MSE) loss of the predicted noised applied to the frequency-time representation of the spectrum. 
We trained for 200 epochs and kept the model optimizing the validation loss.

\section{Experiments and Results}
After inpainting the frequency-time representation, we obtain the reconstructed spectrum with a 1D Fourier Transform on the evolution time axis. All experiments and benchmark metrics are based on the quality of this final reconstructed spectrum since this is the output analyzed by NMR users.
We simulate the partial acquisition by masking rows of the frequency-time representation, which is a fair approximation applied in the partial NMR acquisition \cite{qu2020accelerated, karunanithy2021fid, zheng2022fast, yan2024diffnmr, zhan2024fast}. 

\subsection{Metrics}
The evaluation approach balances both a global perspective, assessing the overall agreement between the original and reconstructed spectra, and a local perspective, focusing on the accuracy of spectral peaks, which are essential for compound characterization in NMR spectroscopy. 

\subsubsection{Global Metrics}
Given our goal of developing a method to reliably reconstruct the raw signal outputted by an NMR machine before any expert signal processing. We use the Mean Squared Error (MSE, lower is better) and the coefficient of determination (R2, higher is better) \tb{to globally evaluate the quality of the predictions}. 
MSE is a standard reconstruction metric \cite{hyberts2010poisson} that quantifies the magnitude of prediction errors, ensuring accuracy in capturing critical spectral features like peak intensities and positions. R² complements this by assessing how well the model explains the variance in the data, providing a normalized measure of fit. Together, these metrics ensure a robust evaluation of the method's precision and reliability.

\subsubsection{Peak-focused Metrics}
We also introduce two metrics specifically designed to assess the integrity of spectral peaks: Peak Hallucination Ratio (lower is better) and Missed Peak Ratio (lower is better). 
The Peak Hallucination Ratio, analogous to the False Detection Rate, measures the proportion of peaks detected in the predicted spectrum that do not correspond to peaks in the reference spectrum. 
Conversely, the Missed Peak Ratio, analogous to the False Negative Rate, evaluates the proportion of peaks in the reference spectrum absent in the predictions. 
These metrics are calculated based on peaks identified by a consistent expert system using the same parameters for all spectra and all methods, ensuring fair and unbiased comparisons. 
By incorporating these peak-specific metrics, we address the critical importance of preserving key spectral features essential for downstream analysis.

\subsection{Guided Sampling Parameters tuning}
\begin{figure}[tb]
    \centering
    \includegraphics[width=\linewidth]{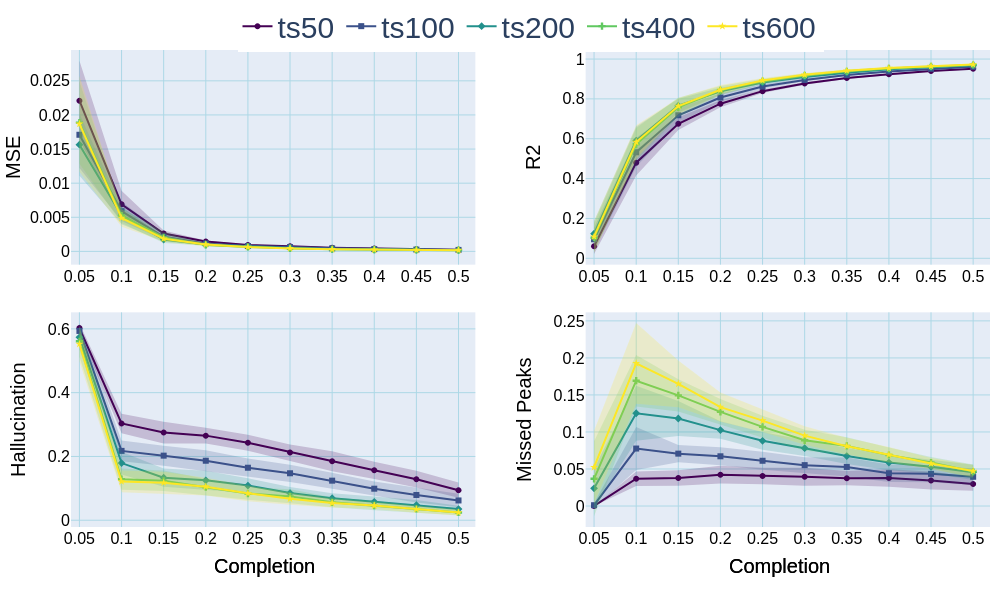}
    \vspace{-0.6cm}
    \caption{Increasing the number of denoising timesteps (ts) improves the global metrics (MSE, R2) and lowers the hallucination ratio but increases the missed peaks.}
    \label{fig:xp-ts}
\end{figure}

The Guided Sampling process depends on several parameters, some being linked to the diffusion model inference and some others to the sampling algorithm. 
We identified the main important parameters to be the number of denoising timesteps and the proportion per step. 
In the following experiments, we always put aside the data point at 95\% masking (i.e., 5\% acquisition completion
), which is the state initialized at random and does not reflect the performances of the guided sampling. 
We optimize these parameters using a greedy approach while keeping all other parameters fixed.

\subsubsection{Percentage of Masking and Percentage of Completion}
Throughout this work, the term `masking proportion' refers to the fraction of data still not acquired, defined as one minus the acquisition completion. For instance, 90\% masking indicates 10\% acquisition completion. 
Metric plots are interpreted from right to left, starting at 100\% masking (0\% completion) and progressing iteratively. In \tb{the} Guided Sampling, each iteration updates the mask by revealing a fixed additional proportion of rows (e.g., 70\% masking is derived from 75\% by completing an additional 5\%).

\subsubsection{Number of denoising timesteps} 
\label{xp:ts}
Diffusion model inpainting is an iterative process that performs several denoising timesteps to obtain the final representation (see Fig.~\ref{fig:xp-ts}). 
We performed a sensitivity analysis on the number of time steps to understand their impact on the Guided Sampling performances.
We observed that the timestep number has a strong effect on the spectrum reconstruction. Increasing the number of time steps leads to better results of the global metrics and hallucination ratio on average after the first guiding step.

The missed peaks ratio, however, seems to be negatively impacted by this number, which we attribute to the fact that a higher timestep number is associated with more accurate uncertainty estimation. The model gains in precision but loses in terms of exploration capacities. 

\subsubsection{Percentage of Completion Per Step} 
\label{xp:mg}

\begin{figure}[tb]
    \centering
    \includegraphics[width=\linewidth]{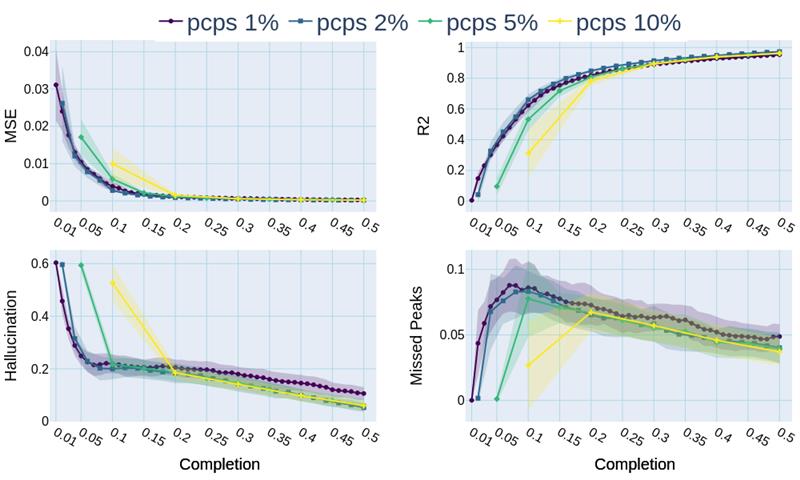}
    \vspace{-0.6cm}
    \caption{An iteration step of 2\% (pcps 2\%) optimizes the global reconstruction metrics and the hallucination peaks at all masking levels.}
    \label{fig:xp-ppsg}
\end{figure}

In this experiment, we observe the impact of the percent of the spectrum acquired at each step (pcps), corresponding to the proportion of evolution times scanned by the NMR machine at each iteration of the guided sampling. 

Shown in Fig.~\ref{fig:xp-ppsg}, after 20\% completion, the iteration step (pcps) 
parameter seems to have a marginal impact on the global reconstruction metrics (MSE and R2), which tend to converge to the same levels, except pcps 1\%, which exhibits worse peak-related metrics overall. 

However, smaller pcps values unlock better performances at high masking levels by leveraging earlier uncertainty guidance. For instance, at 10\% completion, pcps 2\% achieves a 2x lower MSE than pcps 5\%, which in turn has a 2x lower MSE than pcps 10\%.

Another strategy of pcps setting is to use an evolutive percentage. We experimented with gradually increasing percentages (increments by power of 2 and Fibonacci sequence) but did not achieve better results with these approaches. We still believe this strategy is promising but leave this study for future work. 
We also tried to mix random and guided evolution times in varying proportions (80\% guided / 20\% random and 50\% / 50\%) but we could not find any improvements in any metrics. 

In our current context of fixed pcps value, we choose a pcps value of 2\% (pcps 2\%), associated with marginally better global metrics, arguably within the standard deviation range. 
We are further comforted in this choice by illustrating uncertainty distribution, which consistently favors pcps 2\% (c.f., Appendix Fig.~\ref{fig:uncertainty_ppsg}).

\subsection{Baselines}
To show the full capabilities of the Guided Sampling approach and based on the number of time step
experiment from Sec.\ref{xp:ts}, we consider two versions of our Guided Sampling with timestep numbers 200 and 400, denoted as GS200 and GS400, respectively.

We compare our versions with two other sampling methods from the literature, Uniform random sampling, which samples evolution time from a uniform distribution, and Poisson-Gap sampling \cite{hyberts2010poisson}. 
This last method has been established as a state-of-the-art sampling method regarding reconstruction quality. It operates by sampling gaps between evolution time from a Poisson-Gap distribution weighted with a sinus function that gives more importance to lower frequencies.

The sampling approaches are compared based on the reconstructed spectrum quality after inpainting the associated frequency-time domain with our pre-trained diffusion model. Since the model is pre-trained on a denoising task only, the reconstruction performances are not biased by the mask design strategy which allows a fair comparison of the sampling strategies.

\section{Discussion}

\begin{figure}[tb]
    \centering
    \includegraphics[width=\linewidth]{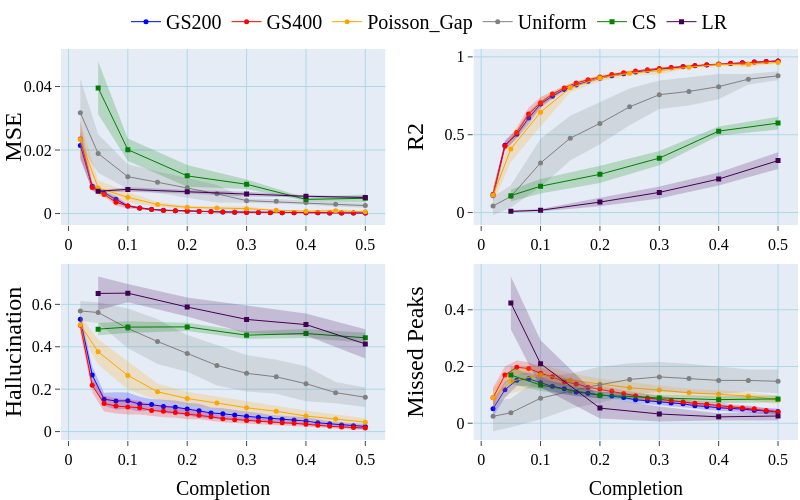}
    \vspace{-0.6cm}
    \caption{Comparison from 2\% to 50\% of acquisition completion. Cross-sampling comparison: Guided Sampling approach (GS) compared with uniform sampling and Poisson-Gap. Cross-model comparison: GS compared with Compressed Sensing and Low-Rank Approximation.
    GS200 and GS400 correspond to the same Guided Sampling strategy with 200 and 400 inference time steps, respectively.}
    \label{fig:final-results}
\end{figure}
In this section, we analyze the spectra reconstructed from the independent test set, focusing on two key aspects: the comparison of sampling strategies (i.e., Guided Sampling, Poisson-Gap Sampling, Uniform Sampling) and the comparison of inpainting methods (i.e., Diffusion model, Compressed Sensing \cite{kazimierczuk2011accelerated}, Low-Rank Approximation \cite{qu2015accelerated}). Together, these comparisons provide a comprehensive understanding of the strengths of our approach in improving NMR spectrum spectroscopy.

\begin{table*}[tb]
\centering
\caption{Comparison when information is scarce (10\% of acquisition completion). Arrows indicate whether higher ($\uparrow$) or lower ($\downarrow$) values are better. The best values are in bold. Halluc. = Hallucination Ratio; M.Peaks = Missed Peaks Ratio.}
\begin{tabular}{l|cc|cc||cc}
\hline
\textbf{Metric} & \textbf{GS200} & \textbf{GS400} & \textbf{Poisson-Gap} & \textbf{Uniform} & \textbf{CS} & \textbf{LR} \\
\hline
MSE $\downarrow$ & \small{\textbf{2.4e$^{-3}$} $\pm$ 3.2e$^{-4}$} & \small{\textbf{2.4e$^{-3}$} $\pm$ 8.1e$^{-4}$} & \small{5.1e$^{-3}$ $\pm$ 1.7e$^{-3}$} & \small{0.01 $\pm$ 3.9e$^{-3}$} & \small{0.02 $\pm$ 3.5e$^{-3}$} & \small{7.6e$^{-3}$ $\pm$ 7.2e$^{-4}$} \\
R$^2$ $\uparrow$ & 0.70 $\pm$ 0.02 & \textbf{0.71} $\pm$ 0.04 & 0.64 $\pm$ 0.09 & 0.32 $\pm$ 0.15 & 0.17 $\pm$ 0.05 & 0.01 $\pm$ 8.6e$^{-3}$ \\
Halluc. $\downarrow$ & 0.15 $\pm$ 0.04 & \textbf{0.12} $\pm$ 0.04 & 0.27 $\pm$ 0.07 & 0.49 $\pm$ 0.09 & 0.49 $\pm$ 0.03 & 0.65 $\pm$ 0.04 \\
M.Peaks $\downarrow$ & 0.14 $\pm$ 0.02 & 0.18 $\pm$ 0.02 & 0.17 $\pm$ 0.04 & \textbf{0.09} $\pm$ 0.07 & 0.13 $\pm$ 0.02 & 0.21 $\pm$ 0.08 \\
\hline
\end{tabular}
\label{tab:metrics_comparison}
\end{table*}

\begin{figure*}[tb]
    \centering
    \includegraphics[width=1\linewidth]{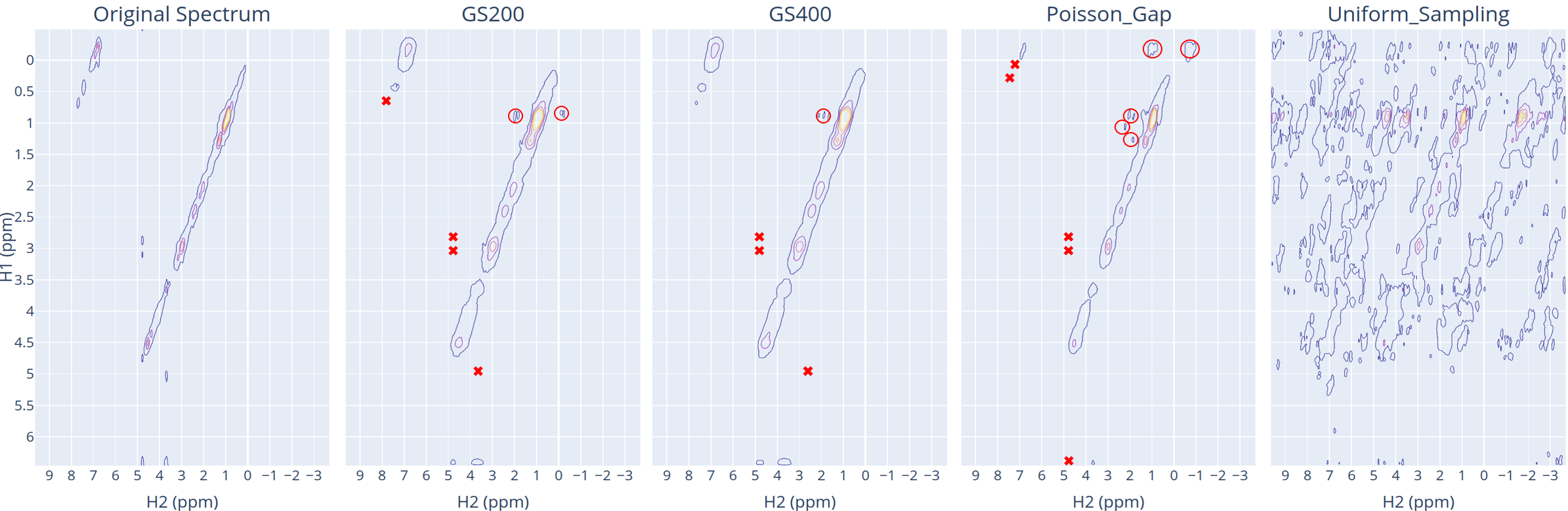}
    \caption{Contours of the CCNOESY NMR spectrum of the phycobilisome linker polypeptide domain of CpcC - 2L8V - nuclei H1 H2; reconstructed from 10\% of pulse sequences. Circles outline hallucinated peaks, and red crosses outline missed peaks. Circles and crosses are omitted from the uniform sampling contour to maintain readability.}
    \label{fig:contours_2l8v}
\end{figure*}

\subsection{Cross-sampling and cross-model comparisons}
As shown in Fig.~\ref{fig:final-results}, the evaluation is based on acquisition completion from 2\% to 50\% (i.e., from 98\% to 50\% of masking).

\textbf{Cross-sampling comparison.}
Within the same diffusion model framework, different sampling strategies show clear differences in performance. Guided Sampling (GS200 and GS400) consistently achieves the best results across all metrics, underscoring the advantages of our proposed sampling strategy. Generally, GS400 slightly outperforms GS200 due to more inference time steps, suggesting a trade-off between computational efficiency and precision.
Poisson-Gap performs comparably in MSE and R$^2$ after 20\% of acquisition completion. 
For the peak reconstruction, Poisson-Gap generates more hallucinated peaks and omits more peaks (Missed Peaks) than Guided Sampling.
However, there is always a performance gap between Uniform Sampling and all NUS strategies (GS and Poisson-Gap).

Note that in terms of Missed Peaks, at low acquisition completion, the best one seems to be uniform sampling, which is actually misleading, as its noise inflates peak counts artificially. Among the other methods, GS200 consistently outperforms, while GS400 is slightly worse than Poisson-Gap for acquisition completion lower than 10\%. 
After 10\% of acquisition completion, GS200 and GS400 follow the same trend and converge.

\textbf{Cross-model comparison.}
While CS \cite{kazimierczuk2011accelerated} and LR \cite{qu2015accelerated} use Non-Uniform Sampling, their reconstruction relies on specialized algorithms that fall short of the diffusion model (Guided Sampling, Poisson-Gap, and Uniform Sampling) in terms of MSE, R$^2$ and Hallucination Ratio.
LR reconstructs fewer Missed Peaks than the diffusion model after 15\% of acquisition completion. As aforementioned, this is due to noise being misidentified as peaks.

\subsection{Performance with scarce information}
Table \ref{tab:metrics_comparison} presents the metrics at 10\% of acquisition completion (i.e., 90\% of the information is unknown.)
GS200 and GS400 achieve an MSE of 2.4e$^{-3}$, which represents a \textbf{52.9\%} reduction compared to Poisson-Gap strategy (5.1$e^{-3}$). Guided Sampling also achieves the highest R$^2$ values ($>$ 0.7), indicating high accuracy and correlation with the reference data.
GS400 achieves the lowest Hallucination Ratio (0.12), a \textbf{55.6\%} reduction compared to Poisson-Gap (0.27), followed by GS200 (0.15). Both methods effectively suppress false peaks, making them reliable for practical use.
The diffusion model with uniform sampling, Compressed Sensing, and Low-Rank Approximation perform the worst.

\textbf{Implications of standard deviations}
Guided Sampling strategy also exhibits greater stability. At 10\% of acquisition completion, GS200 has relatively low variability across all metrics, particularly in MSE (3.2e$^{-4}$), ensuring consistent performance. GS400, while slightly more variable, remains reliable.
Poisson-Gap shows moderate variability, while the Uniform sampling, CS, and LR exhibit higher standard deviations, reflecting its instability and inconsistent results.

\begin{figure}[tb]
    \centering
    \includegraphics[width=\linewidth]{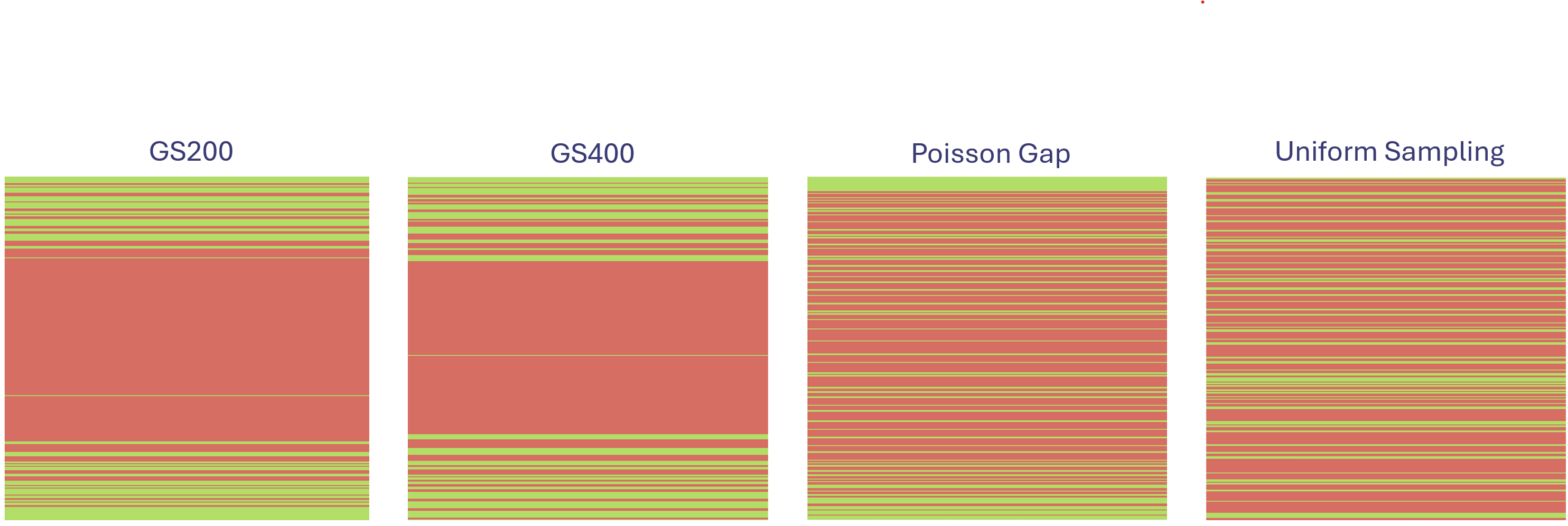}
    \caption{Masks obtained after 30\% sampling of 2L8V-CCNOESY NMR acquisition in Frequency-Time domain. Red indicates rows not yet sampled.}
    \label{fig:mask_70_2l8v}
\end{figure}
\begin{figure}[tb]
    \centering
    \includegraphics[width=\linewidth]{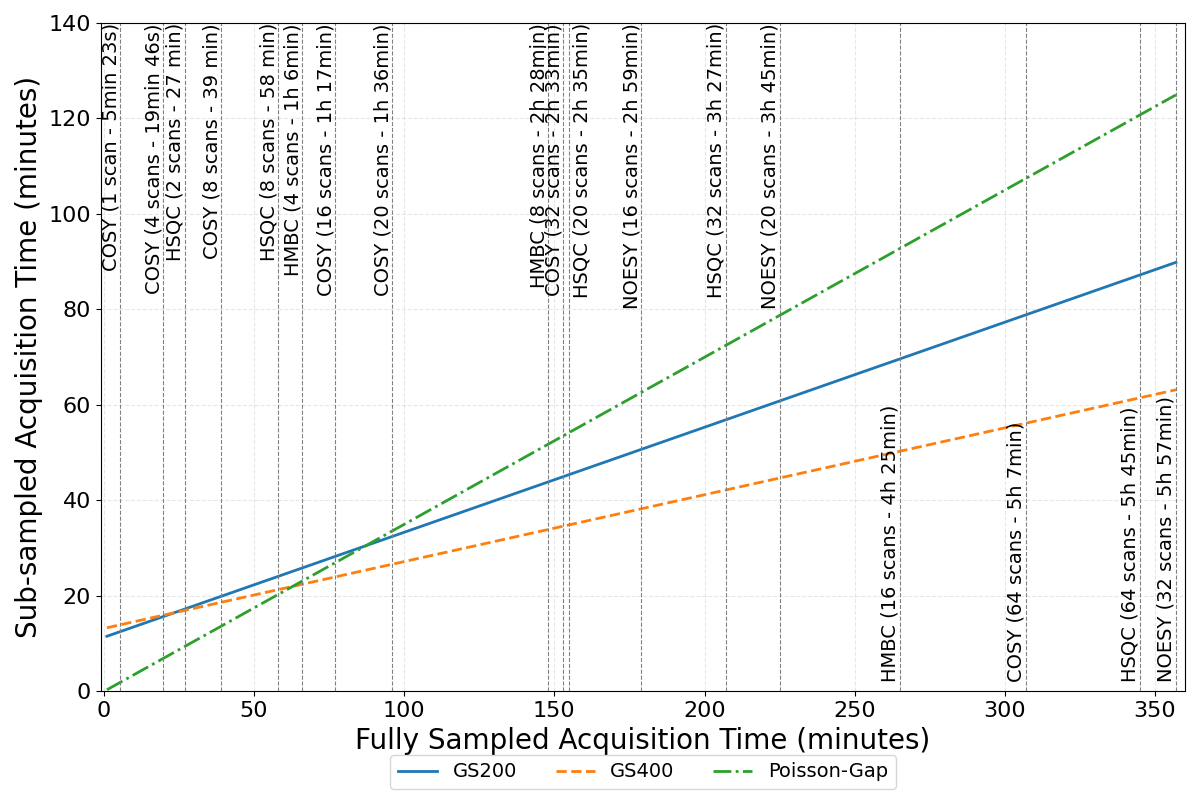}
    \caption{Wall time (minutes) required by sub-sampling strategies to achieve a hallucination ratio below 10\% as a function of fully sampled acquisition time.}
    \label{fig:wall_time}
\end{figure}
\subsection{Case study}

Applied to the CCNOESY NMR spectrum of the phycobilisome linker polypeptide domain of CpcC - 2L8V - nuclei H1 H2 \cite{2l8v}, the reconstructed results (displayed in Fig.~\ref{fig:contours_2l8v}) show the practical effects of the method choice. 
After collecting 10\% of the pulse sequences, Uniform sampling, CS and LR show a high proportion of noise, which severely compromise the peak analysis. 
Poisson-Gap strongly improves the signal reconstruction and retains most of the 
signal information, but contains a small amount of missed peak and hallucinations. The Guided Sampling approach (especially GS400) contains the least hallucinations and recovers most of the signal.

Interestingly, the Guided Sampling mask in the frequency-time domain resembles the Poisson-Gap mask, with significant weight distributed at long and short evolution times (Fig.~\ref{fig:mask_70_2l8v}). However, Guided Sampling amplifies this pattern, leaving nearly no rows sampled in the central region of the mask (medium evolution times). This highlights that most information needed to reconstruct the spectrum is concentrated at the edges of the frequency-time domain.

\subsection{Wall Time Comparison}
All training and inference were conducted on an Nvidia H100 GPU (80 GB memory). Guided sampling steps, including sample generation and uncertainty prediction, took 1 min 8 s for GS200 and 2 min 12 s for GS400.

Using the average wall times for fully sampled 2D NMR acquisitions \cite{nmr_xp_time}, we estimated the total wall time (diffusion model inference + pulse sequence acquisition) required for each NUS strategy to achieve a hallucination ratio below 10\% (Fig.~\ref{fig:wall_time}).

GS200 and GS400 consistently outperformed Poisson-Gap sampling in experiments where fully sampled acquisition times exceeded 87 and 62 minutes, respectively. These time advantages covered most experiments with more than eight scans. Compared to Poisson-Gap, GS200 achieved time reductions of \textbf{10.2\%}, \textbf{19.2\%}, and \textbf{23.7\%}, while GS400 achieved \textbf{28.6\%}, \textbf{39.1\%}, and \textbf{44.3\%} for 2-hour, 3-hour, and 4-hour experiments, respectively.

These time savings can be further enhanced by optimizing GPU utilization. For example, running two NMR acquisitions in parallel would allow the GPU to process the guiding step of one experiment while the NMR machine acquires data for another, effectively removing GPU time from the total wall time. Alternatively, sample reconstructions for uncertainty estimation could be distributed across multiple GPUs.

\section{Conclusion}
In this work, we introduced diffusion models to NMR spectroscopy for the first time, demonstrating their ability to reconstruct sub-sampled 2D spectra with reconstruction 
performance surpassing state-of-the-art methods. We also proposed a novel adaptive sampling strategy that leverages the uncertainty of the diffusion model to dynamically prioritize sampling regions, achieving more efficient sampling while maintaining superior reconstruction quality.

Guided Sampling delivered better reconstruction metrics with fewer hallucinated and missed peaks compared to existing methods. By adjusting the denoising timestep parameter, users can balance the trade-off between computational efficiency and precision, tailoring the method to specific experimental needs. 
Additionally, Guided Sampling exhibited lower variability across metrics, highlighting its stability and reliability. We also analyzed acquisition time as 
a function of the duration of the full sample scan, showing clear benefits for experiments lasting more than 62 minutes or involving more than 8 scans.

This approach is highly promising for higher-dimensional spectra, where acquisition times increase exponentially with dimensions. Furthermore, it opens new avenues for refined sampling strategies based on uncertainty distribution analysis, such as prioritizing extreme evolution times. 
Finally, diffusion models hold significant potential for other NMR optimizations, 
including artificial resolution enhancement and efficient spectrum-to-structure matching.



\section*{Impact Statement}
This paper presents work whose goal is to advance the field of 
Machine Learning. There are many potential societal consequences 
of our work, none which we feel must be specifically highlighted here.

\newpage
\appendix
\section{Appendix}
\subsection{Row-wise Uncertainty with different pcps}
\begin{figure}[h]
    \centering
    \includegraphics[width=\linewidth]{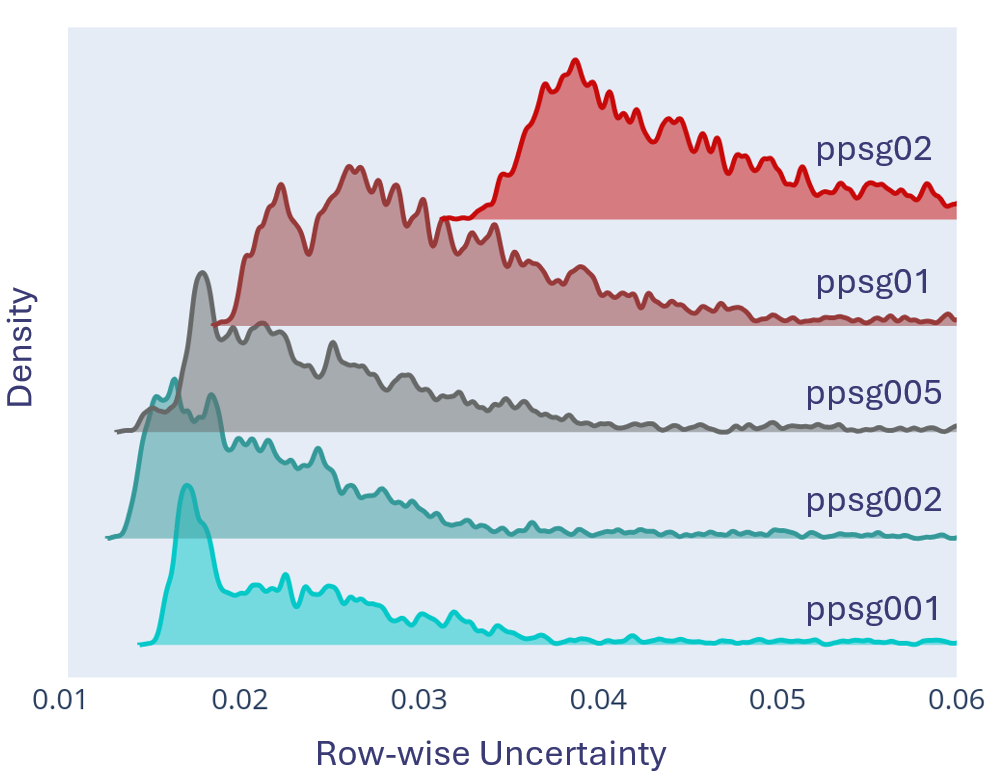}
    \caption{Small acquisition steps (i.e., pcps values) are associated with lower uncertainty distribution (here, at 75\% masking), with an optimum at 2\% (pcps 2\%).}
    \label{fig:uncertainty_ppsg}
\end{figure}

We illustrate the Row-wise Uncertainty with different pcps in Fig.~\ref{fig:uncertainty_ppsg}. The x-axis represents the row-wise uncertainty. If the distribution is closer to the left, it means that the uncertainty is lower.

\subsection{Wall Time Comparison}
We add the comparison for the wall time on the missed peaks ratio below 10\%.
The gains are even more pronounced for missed peak ratios than for hallucination ratios. In this case, GS200 is more efficient than other methods for experiments exceeding 39 minutes. For 1-hour, 2-hour and 3 hour experiments, GS200 is respectively \textbf{20\%}, \textbf{35\%}, and \textbf{50\%}, faster than Poisson-Gap sampling.
\begin{figure}[h]
    \centering
    \includegraphics[width=1\linewidth]{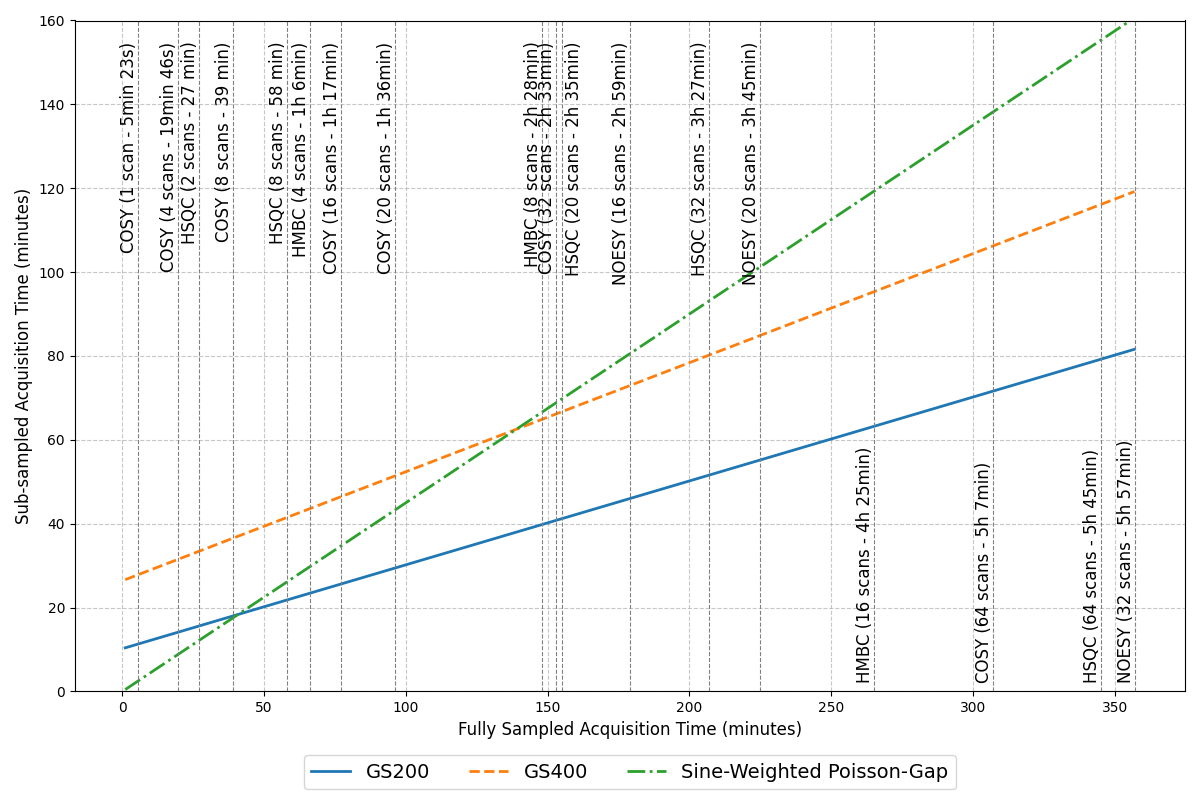}
    \caption{Wall time (in minutes) required by the sub-sampling methods to reach a missed peaks ratio below 10\%, as a function of fully sampled acquisition time.}
    \label{fig:iso_missed_peaks_at_10pct}
\end{figure}
\bibliographystyle{unsrt}  
\bibliography{references}

\end{document}